\documentclass[conference]{IEEEtran}
\IEEEoverridecommandlockouts
\usepackage{cite}
\usepackage{amsmath,amssymb,amsfonts}
\usepackage{algorithm}
\usepackage{algorithmicx}
\usepackage[noend]{algpseudocode}
\usepackage[caption=false]{subfig}

\usepackage{graphicx}
\usepackage{textcomp}
\usepackage{xcolor}
\usepackage{tcolorbox}
\usepackage{graphicx}

\usepackage{hyperref}

\usepackage{comment}
\usepackage{todonotes}

\usepackage{latexsym}
\usepackage{subfig}
\usepackage{float}

\usepackage{colortbl}
\usepackage{multirow}
\usepackage{adjustbox}

\usepackage{chngcntr}
\usepackage{booktabs}
\usepackage{tabularx}
\usepackage{caption}

\usepackage{hyperref}

\def\BibTeX{{\rm B\kern-.05em{\sc i\kern-.025em b}\kern-.08em
    T\kern-.1667em\lower.7ex\hbox{E}\kern-.125emX}}

\newtheorem{definition}{Definition}

\renewcommand{\arraystretch}{1.5}
\definecolor{lightgreen}{rgb}{0.8, 1, 0.8}
\definecolor{lightred}{rgb}{1, 0.8, 0.8}
\definecolor{lightblue}{rgb}{0.8, 0.85, 1}

\begin{document}
\title{Discovery and Simulation of Data-Aware Business Processes\thanks{Work funded by European Research Council (PIX project) and Estonian Ministry of Education \& Research via Estonian Centre of Excellence in AI.}
}
%
%

\author{\IEEEauthorblockN{Orlenys L\'opez-Pintado }
\IEEEauthorblockA{\textit{University of Tartu} \\
Tartu, Estonia \\
orlenyslp@ut.ee}
\and
\IEEEauthorblockN{Serhii Murashko}
\IEEEauthorblockA{\textit{University of Tartu} \\
Tartu, Estonia \\
serhii.murashko@ut.ee}
\and
\IEEEauthorblockN{Marlon Dumas}
\IEEEauthorblockA{\textit{University of Tartu} \\
Tartu, Estonia \\
marlon.dumas@ut.ee}
}

\maketitle              
\graphicspath{{figs/}}
\begin{abstract}
Simulation is a common approach to predict the effect of business process changes on quantitative performance. The starting point of Business Process Simulation (BPS) is a process model enriched with simulation parameters. To cope with the typically large parameter spaces of BPS models, several methods have been proposed to automatically discover BPS models from event logs. Virtually all these approaches neglect the data perspective of business processes. Yet, the data attributes manipulated by a business process often determine which activities are performed, how many times, and when. This paper addresses this gap by introducing a data-aware BPS modeling approach and a method to discover data-aware BPS models from event logs. The BPS modeling approach supports three types of data attributes (global, case-level, and event-level) as well as deterministic and stochastic attribute update rules and data-aware branching conditions. An empirical evaluation shows that the proposed method accurately discovers the type of each data attribute and its associated update rules, and that the resulting BPS models more closely replicate the process execution control flow relative to data-unaware BPS models.


\end{abstract}


\section{Introduction}
\label{sec:introduction}

Business Process Simulation (BPS) is a technique to predict how one or more changes to a process will affect its performance~\cite{RozinatMSA09}. It allows analysts to answer ``what-if" questions such as ({\bf S1}) how would a shift in consumer preferences from one product category to another affect the performance of an order-to-cash process; or ({\bf S2}) what would be the effect of changes in the distribution of loan amounts on the cycle time of a loan origination process? The starting point of BPS is a process model enriched with simulation parameters (a BPS model). A simulation engine interprets the BPS model, yielding a simulated event log and performance metrics.


BPS models consist of numerous elements and parameters that need to be carefully tuned to reflect the behavior of the modeled business process. Several studies~\cite{RozinatMSA09, CamargoDG20, Estrada-TorresC21, Lopez-PintadoDB24} have proposed data-driven simulation methods, which automatically discover BPS models from event logs.
These methods focus on discovering BPS models along the control flow, organizational, and temporal perspectives. While some existing methods discover conditions on data attributes or firing rules based on data variables, none is able to discover how data attributes are initialized and updated during the execution of cases in a business process. Hence, the resulting BPS models cannot be used to answer what-if questions in the style of ({\bf S1}) and ({\bf S2}).

This paper addresses the outlined limitations by proposing (1) a data-aware business process simulation approach supporting different types of data attributes alongside data initialization (a.k.a., generation) and update rules and branching conditions based on data attributes and (2) a method to discover such data-aware simulation model from event logs.\footnote{In this paper, ``data-driven simulation'' refers to simulation methods based on BPS models discovered from event logs, whereas ``data-aware simulation model'' refers to a simulation model that incorporates information on how data items are produced and used, and ``data attributes'' refers to a mechanism to capture data in a simulation model.} The paper presents an empirical evaluation showing that the proposed method can accurately classify data attributes and discover their associated update rules. Besides, it shows that data-aware simulation models with branching conditions discovered from event logs replicate process control flow more accurately than (or comparable to) data-unaware models.


The rest of the paper is structured as follows. Section~\ref{sec:related} discusses related work. Section~\ref{sec:simulation-model} describes and formalizes the data-aware simulation approach. Section~\ref{sec:approach}  proposes the corresponding method to discover the simulation models. Section~\ref{sec:evaluation} empirically evaluates the proposal, and Section~\ref{sec:conclusion} concludes and sketches future work.
\section{Related Work}
\label{sec:related}


Rozinat et al.~\cite{RozinatMSA09} propose a method to discover a BPS model from an event log. The method first discovers a control-flow model using automated process discovery. It then enhances it with activity durations and roles. Roles are discovered using clustering so that resources that usually perform the same activities are grouped in one role. Finally, the model is enhanced with branching conditions discovered via decision tree learning. However, this approach does not capture how the values of the data attributes used in the conditions are initialized and updated. The present paper fills this gap. Also, the present paper differs from~\cite{RozinatMSA09} in that it combines branching conditions with branching probabilities so that probabilities are used when the accuracy of the conditions is low.


Camargo et al.~\cite{CamargoDG20} extend the approach of Rozinat et al. in three ways. First, it lifts the assumption that activity durations always follow a normal distribution. Second, it discovers branching probabilities instead of branching conditions, thus ensuring that the resulting model is executable (without manually adding data attribute initialization and updating rules). Finally, it adds a hyperparameter optimization phase to fine-tune the parameters used to discover the control-flow model, the roles, and every other component of the BPS model. The use of hyperparameter optimization consistently improves the accuracy of the resulting BPS model relative to the approach of Rozinat et al. On the other hand, the method of Camargo et al. does not discover data attributes or conditions.

Estrada-Torres et al.~\cite{Estrada-TorresC21} extend the method of Camargo et al. to capture resource availability calendars and multitasking behavior, further enhancing the accuracy of the resulting models.\footnote{In this paper, we do not handle multitasking as this is an orthogonal issue.} Later, López-Pintado et al.~\cite{Lopez-PintadoDB24} refine the approach of Estrada-Torres et al. (and by transitivity that of Camargo et al. and Rozinat et al. to lift the restrictions that (i) roles must be disjoint; (ii) all resources of a role must have the same calendar; and (iii) an activity can only be associated to one role. 
The authors show that this \emph{differentiated resources} approach outperforms the method by Estrada-Torres et al. and by transitivity those of Camargo et al. and Rozinat et al. Accordingly, we use this latter approach as a baseline.


Pourbafrani et al.~\cite{PourbafraniA22} discover system dynamics simulation models using process mining, focusing on high-level aggregated behaviors with causal-loop and stock-flow diagrams. 
Other recent approaches combine BPMN simulation modeling with deep learning models to generate timestamped event sequences~\cite{CamargoDR22} and to predict waiting and processing times\cite{MeneghelloFG23}, but without considering data attributes.

De Leoni et al.~\cite{LeoniVLM23}, building upon previous studies~\cite{BurkeLW20, MannhardtLSL23}, discover stochastic Petri nets extended with a notion of ``process state'' based on ``variables''. The approach computes transition weights through logistic regression based on these states. In contrast, the approach presented in this paper uses data attributes of different types, update rules, and data-aware branching conditions to model the data perspective. This explicit support for data attributes and branching conditions allows users to define simulation scenarios involving changes in data distributions, data updates, and decision logic.

Decision mining techniques~\cite{MannhardtLRA16, SmedtHBV19}  identify decision points within a process and discover data-aware branching conditions for each path. In contrast, the proposed data-aware process simulation model integrates decision conditions into a broader approach, categorizing data attributes and using deterministic and stochastic update rules.

\section{Data-Aware Process Simulation Models}
\label{sec:simulation-model}

Our notion of Data-Aware BP Simulation Model (DAS), formalized in Definition~\ref{def:das} integrates standard Business Process Model and Notation (BPMN) elements enhanced with simulation parameters. The BPMN model captures the control-flow perspective, i.e., the process activities and their relations. On top of that, the simulation parameters target the organizational, temporal, and data dimensions. Although this paper focuses on discovering and representing data components and decision-making mechanisms, Definition~\ref{def:das} also introduces the notions of resource allocation and inter-arrival case schemas required for modeling the organizational and temporal perspectives as part of the simulation environment. 

\begin{definition}
\label{def:das}
A Data-Aware BP Simulation Model (DAS) builds upon standard BPMN elements \( \langle E, A, G, F \rangle \) – with events (E), activities (A), gateways (G), and flow arcs (F) – and enhanced with specialized simulation parameters as follows: 

\begin{itemize}

    \item \( RS = \langle \text{TRMap}, \text{ProcTimes}, \text{RAvail} \rangle \) is the Resource Allocation Schema. This schema includes \text{TRMap}, mapping of resources to tasks; \text{ProcTimes}, which are functions modeling the time required for resources to complete assigned activities; and \text{RAvail}, representing the availability calendars of resources for activity execution.

    \item \( AT_{AC}: \mathcal{P}(\mathbb{R}^+) \times {\tt Intervals} \) models case creation times by combining \( AT \), the probability density function for inter-arrival times, with \( AC \), the calendar in which new cases can be created.
    
    \item \( D_g = \{g_1, g_2, \ldots, g_m\} \) is a finite set of global attributes.

    \item \( D_c = \{c_1, c_2, \ldots, c_k\} \) is a finite set of case attributes.

    \item \( D_e = \{e_1, e_2, \ldots, e_l\} \) is a finite set of event attributes. 

    \item \( UR = \{ur_1, ur_2, \ldots, ur_n\} \) is a set of update rules \( ur_i: (D_g \cup D_c \cup D_e) \times \mathcal{S} \rightarrow \mathcal{S} \). Each \( ur_i \in UR \) consists of a combination of functions and conditional logic to dynamically modify the current data state ($s \in \mathcal{S}$) of any global, case, or event attributes, resulting in a new data state in the domain of possible data states $\mathcal{S}$.

    \item \( MUR: UR \rightarrow (A \cup \{{\tt Case\_Creation}\})   \) maps each update rule \( ur \in UR \) to the step in the process where this rule is executed, which may be when the case is created or when a task is completed.
    
    \item \( BC: G \times (\mathcal{S}_{D_g} \times \mathcal{S}_{D_c} \times \mathcal{S}_{D_e}) \rightarrow F \times \{ True, False \} \) is a function that, given a gateway and a data state of the process, assigns a truth value to each flow $F$, such that flows $f \in F$ s.t., $source(f) \neq G$ are assigned a $False$ value, while flows $f \in F$ s.t., $source(f) = G$ may be $False$ or $True$.
    
    \item \( BP: F \rightarrow [0..1] \) is a function that assigns a probability to each flow, such that flows whose source is not a gateway (unconditional flows in the BPMN terminology) get a value of 1, while flows whose source is a gateway (conditional flow) may get a value between zero and one.    
\end{itemize}
\end{definition}

In Definition~\ref{def:das}, attributes are partitioned into sets $D_g$, $D_c$, and $D_e$ based on their scope (local to a process instance vs global across all cases of a process) and their mutability (static or dynamic). A global attribute (e.g., {\bf\it available-hospital-beds}) has a  shared value across all process cases. They can capture critical sections, e.g., to prevent new patient admissions when the hospital reaches its capacity. A case attribute (e.g., {\bf\it reason-for-admission}) may take a different value for each case. A case attribute is static -- its value is determined at the start of a case and cannot be updated. An event attribute (e.g., {\bf\it assigned-doctor} or {\bf\it diagnostic-test-results}) has a local scope, and it is dynamic -- an update rule may change its value.

An update rule changes the value of a global or an event attribute during the execution of a case. For example, when a patient is assigned a bed, the case attribute {\bf\it available-hospital-beds} is decremented. This behavior may be captured by an update rule associated with a task ``Assign bed''. Similarly, an update rule associated with a task ``Record diagnostic'' may change the value of an event attribute {\bf\it diagnostic-test-results}. An update rule may be used to initialize a case attribute, e.g., to assign a value to the {\bf\it reason-for-admission} attribute. Conceptually, such update rules are executed when a case is created. The mapping $MUR$ determines whether an update rule is fired upon case creation or when a task instance is completed (e.g., after ``Assign bed'').

In Definition~\ref{def:das}, the ``data state'' refers to the set of ``attribute: value'' pairs (of all the defined attributes in the process model) at a given point in the process simulation. This state is altered by executing update rules as the simulation unfolds.



The data state of the process determines which flows may be taken when the control-flow state of the simulation reaches a decision gateway. In Definition~\ref{def:das}, this is captured via function \textsf{BC} (standing for Branching Conditions), which assigns a truth value to each outgoing flow of a decision gateway based on the data state. This function is defined at an abstract level, but concretely, it is a boolean expression over the data attributes.

Definition~\ref{def:das} additionally supports the selection of decision gateway branches based on probabilities. This is captured by the function \textsf{BP} (standing for Branching Probabilities), which assigns a probability to each conditional flow.




The execution semantics at each decision gateway follow a two-staged approach. First, the branching conditions determine which flows may be taken. Next, the branching probabilities are used to determine which one(s) is/are selected among the flows that may be taken. Specifically, when a gateway is reached during a case simulation, the boolean function associated with each outgoing conditional flow is evaluated, thus assigning a value of $True$ or $False$ to each flow. Flows are then selected as follows:
\begin{itemize}
\item 
If only one of the flows gets a $True$ value, it is selected. 
\item 
If multiple flows get a $True$ value, the behavior depends on the type of decision gateway:
  \begin{itemize}
    \item For an exclusive (XOR) gateway, the flows that fulfill the branching probabilities of these $True$ flows are summed, then normalized to a total probability mass of 1.0, and, finally, one of the flows is randomly selected based on these probabilities. 
    \item For an inclusive (OR) gateway, a coin flip is made for each flow with a $True$ value considering its associated branching probability. If the coin flips lead to no flow being selected, the default flow is taken. It may be that multiple flows are selected.
  \end{itemize}
\item If no flow gets a $True$ value, the \emph{default flow} is taken, as per the standard BPMN semantics.
\end{itemize}

This approach allows us to define purely stochastic choices by setting the branching conditions at a gateway to ``always $True$'' so that the branching probabilities drive the choice. Conversely, we can define a purely data-driven choice by assigning branching conditions to a gateway and setting its branching probabilities to 1. Hybrids between these two extremes are supported, e.g., a gateway where some branches have branching conditions while others only have probabilities.

    
    



\section{Discovering Data-Aware Simulation Models}
\label{sec:approach}


In the following, we describe our approach to discovering a simulation model (as in Definition~\ref{def:das}) from an event log. An event log is a set of events. Each event represents the execution of an activity instance within a case of a business process.\footnote{This type of event log is also known as an activity instance log.} An event consists of a case identifier, an activity label, a resource identifier, start and end timestamps, and additional attributes capturing inputs or outputs of the activity instance or attributes of the case. Given that a case identifier ID is present in an event log, a trace is the sequence of events with the same ID as a case identifier, chronologically ordered by start timestamp. Thus, a trace captures the execution of a case.

 The proposed method involves three main steps. (1) Extracting and classifying global ($D_g$), event ($D_e$), and case ($D_c$) attributes. (2) Computing the update rules ($UR$) that define how to modify these attributes throughout the simulation and mapping these rules to the respective activities responsible for such updates ($MUR$). (3) Discovering the branching conditions ($HBC = BC \cup BP$) related to the decision points within the process flow. The remaining components in Definition~\ref{def:das}, including control flow, organizational, and temporal dimensions (outside the scope of this paper), were discovered using the approach in~\cite{Lopez-PintadoDB24}.

 \subsection{Data Attribute Classification}


 Attributes in an event log are not explicitly labeled as global, event, or case attributes. This metadata is required for simulation, as each attribute type has different execution semantics. Thus, one of the challenges of discovering a simulation model is identifying each attribute's type. Separating event attributes from global attributes requires an analysis of update patterns across traces, as the same sequence of attribute values could result from global or event-level updates, creating ambiguity in determining the attribute's proper classification. A second challenge is that event logs give us the value of attributes after the activity instance completes. However, we also need the attribute's value when the activity instance starts to discover update rules. Thus, another challenge is to infer such pre-execution values. Finally, given the pairs of values before and after the executions of a given activity, a third challenge is to infer a rule that captures how the post-execution value is derived from the pre-execution one.

 Conversely to global and event attributes, which are subject to ambiguity, identifying case attributes is straightforward because their values remain constant within each process case. To account for potential data noise, we include a threshold parameter specifying the ratio of instances where the invariability condition must hold in the discovery. For example, with a threshold of 0.9, we classify an attribute as a case attribute if its value is constant in 90\% of the cases. The remaining dynamic attributes will be categorized as global or event attributes.


 Given the inherent ambiguity in classifying dynamic attributes as global or event-level in a log, our approach evaluates each attribute as both types to find the best fit using regression analysis. Algorithm~\ref{algo:attr_classif} outlines the general method for classifying attributes and discovering their update rules. Initially, each event is split into two parts based on their starting and ending times. As shown in Algorithm~\ref{algo:split-events}, the starting part of the event does not include any information about data attributes, as these values are initially unknown and thus estimated in the following steps.


\begin{algorithm}[tp]
\begin{algorithmic}[1]
\scriptsize
\Function{ClassifyAsAttribute}{$L$: {\sc EventLog}, $INIT_G$: {\sc DataGenerator}}
\State $L^{SP}$ $\gets$ {\sc SplitEventLog}($L$)
\State \textcolor{red}{--- Attempting data-attribute as a event attribute ---}
\State $D^E$ $\gets$ \{\} \Comment{\textcolor{blue}{Stores data-updates for each pair $<activity, attribute>$}}
\For{{\bf each} {\sc CASE-ID} $cI \in L^{SP}$}
    \State $T$ $\gets$ {\sc FilterEventsByCaseID}($L^{SP}$, $cI$)
    \State {\sc SortByTimestamps}($T$)
    \State {\sc FindDataValues}($T$, $D^E$, $INIT_G$)
\EndFor

\State \textcolor{red}{--- Attempting data-attribute as a global attribute ---}
\State $D^G$ $\gets$ \{\} \Comment{\textcolor{blue}{Stores data-updates for each pair $<activity, attribute>$}}
\State {\sc SortByTimestamps}($L^{SP}$)
\State {\sc FindDataValues}($L^{SP}$, $D^G$, $INIT_G$)

\State \textcolor{red}{--- Selecting best attribute classification ---}
\State $UR$ $\gets$ \{\} {\textcolor{blue}{Stores update rules for each pair $<activity, attribute>$}}
\For{{\bf each} pair  $<$ $\alpha$: Activity, $B$: attribute-type $>$ $\in L$}
    \State $UR^G$ $\gets$ {\sc FindUpdateRule}($D^G$, $\alpha$, $B$) \Comment{\textcolor{blue}{$UR$ as Global Attribute}}
    \State $UR^E$ $\gets$ {\sc FindUpdateRule}($D^E$, $\alpha$, $B$) \Comment{\textcolor{blue}{$UR$ as Event Attribute}}
    \State $UR[\alpha][B]$ $\gets$ {\sc MinErroUR}($L$, $UR^G$, $UR^E$)
\EndFor

\State {\bf return} $UR$
\EndFunction

\end{algorithmic}
\caption{Dynamic Attributes Classification}
\label{algo:attr_classif}
\end{algorithm}

\begin{algorithm}[tp]
\begin{algorithmic}[1]
\scriptsize
\Function{SplitEventLog}{$L$: {\sc EventLog}}
\State $L^{SP} \gets []$
\For{{\bf each} event $e \in L$}
    \State {\sc Add}($L^{SP}$, $<$CASE-ID$[e]$, ACTIVITY$[e]$, START-TIME$[e]$$>$)
    \State {\sc Add}($L^{SP}$, $<$CASE-ID$[e]$, ACTIVITY$[e]$, END-TIME$[e]$, ATTRS$[e]$$>$)
\EndFor
\State {\bf return} $L^{SP}$
\EndFunction

\end{algorithmic}
\caption{Splitting Events by Start and End Times}
\label{algo:split-events}
\end{algorithm}

\begin{algorithm}[tp]
\begin{algorithmic}[1]
\scriptsize
\Function{FindDataValues}{$L^{SP}$: {\sc EvList}, $D$: {\sc Dict}, $INIT_G$: {\sc Generator}}

\For{{\bf each} data-attribute type $B \in L^{SP}$}
    \State $\beta$ $\gets$ $INIT_G[B]$ \Comment{\textcolor{blue}{Initializing data-attribute value from its generator}}
    \State $A'$ $\gets$ \{\} \Comment{\textcolor{blue}{Dictionary of $<activity: data-value>$}}
    \For{{\bf each} event $e \in L^{SP}$}
        \State $\alpha$ $\gets$ {\sc ACTIVITY}[$e$]
        \If{{\sc IS-START}[$e$]}
            \State $A'$[$\alpha$] $\gets$ $\beta$
        \EndIf
        \If{{\sc IS-END}[$e$]}
            \State $\beta$ $\gets$ {\sc ATTRIBUTE-VALUE}[$B$, $e$]
            \State $D$[$\alpha$][$B$].{\sc Append}($<A'$[$\alpha$], $\beta>$)
        \EndIf
    \EndFor
\EndFor
\EndFunction

\end{algorithmic}
\caption{Estimating Activity Starting Data Values}
\label{algo:estimate_start}
\end{algorithm}


 Next, Algorithm~\ref{algo:attr_classif} estimates the data values at the beginning of each activity in the log. Those values may vary depending on whether it is a global or event attribute. Lines 3-8 attempt each attribute as an event attribute, i.e., handled by each process case independently. To this end, it sorts events within each case by their timestamps, resets attribute values to their default (e.g., 0 if numeric) at the beginning of each case, and updates them based on the most recent completed activity within the same case. This approach ensures that event attribute candidates reflect each case's specific context and progression without being influenced by other cases.

 For attempting attributes as global (lines 9-12), Algorithm~\ref{algo:attr_classif} does not reset values between cases but treats the entire log as a continuous sequence of events. It globally sorts all events by their timestamps, so the value of a global attribute at the start of an activity depends on the most recent completed activity in the entire log. This method captures cumulative changes across all cases, ensuring that global attributes maintain a continuous and cumulative value throughout the process.


 Algorithm~\ref{algo:estimate_start}, invoked in lines 8 and 12 of Algorithm~\ref{algo:attr_classif}, respectively, under event and global attribute assumptions, estimates starting data values for each pair activity-attribute ($\alpha$, $B$). Since event logs do not provide initial attribute values, the algorithm uses a predefined data generator ($INIT_G$) to initialize these values. This generator could assign default values, such as 0, for numeric attributes or use other specified functions. As Algorithm~\ref{algo:estimate_start} iterates over each event, it records initial values at start events from the previously completed activity while keeping the values observed in the log at end events, thus tracking attribute states required to estimate the update rules in the following steps. Back to Algorithm~\ref{algo:attr_classif}, in line 13, it already retrieved the entire data flow for each attribute under two possible scenarios, i.e., local ($D^E$) and global ($D^G$). Then, it derives the corresponding update rules and selects, for each attribute, the classification (global/event) that leads to the minimum error (line 18). Specifically, it chooses the classification whose associated update rule most closely matches the observed data flow in the input log.



\subsection{Update Rules Extraction}


 Lines 13-17 of Algorithm~\ref{algo:attr_classif} extract the update rules candidates for each attribute, assuming both global and event classification scenarios. To this end, Algorithm~\ref{algo:update_rules} is invoked to determine the most suitable update rules. Here, attributes from event logs are broadly classified into categorical and continuous (numerical). Categorical attributes represent discrete values or categories, such as the status of an order (``pending,'' ``shipped,'' ``delivered'') or the type of task performed. Continuous attributes, on the other hand, represent numerical values that can be measured on a continuous scale, such as inventory levels or costs. For categorical attributes, update rules involve transitioning between different states, while for continuous attributes, they involve mathematical functions that describe changes in numerical values. Besides, generators are a specific update rule type that produces new values independently of the current attribute state, often based on predefined probabilities or distributions. Thus, unlike typical update rules, which consider the attribute's current value to determine the next value, generators do not consider the current state.

\begin{algorithm}[tp]
\begin{algorithmic}[1]
\scriptsize
\Function{FindUpdateRules}{$D$: {\sc DataTransitionsDict}, $\alpha$, $B$}
\If{{\sc IS-NUMERICAL}[$B$]}
    \State $UR_1$ $\gets$ {\sc LinearRegression}($\forall$ pair $<\beta_0, \beta_n> \in$ $D[\alpha][B]$)
    \State $UR_2$ $\gets$ {\sc DecisionTreeRegression}($\forall$ pair $<\beta_0, \beta_n> \in$ $D[\alpha][B]$)
    \State $UR_3$ $\gets$ {\sc DistributionUR}($\forall$ $\beta_n - \beta_0,$  $<\beta_0, \beta_n> \in$ $D[\alpha][B]$)
    \State $UR_4$ $\gets$ {\sc DistributionGen}($\forall$ $\beta_n \in$ $D[\alpha][B]$)
\EndIf
\If{{\sc IS-CATEGORICAL}[$B$]}
    \State $UR_5$ $\gets$ {\sc MarkovMatrix}( $\forall$ pair $<\beta_0, \beta_n> \in$ $D[\alpha][B]$)
    \State $UR_6$ $\gets$ {\sc ProbabilisticGenerator}( $\forall$ $\beta_n \in$ $D[\alpha][B]$)
\EndIf
\State {\bf return} {\sc MinErrorUR}($D$, $UR_1$, $UR_2$, $UR_3$, $UR_4$, $UR_5$, $UR_6$)

\EndFunction

\end{algorithmic}
\caption{Update Rules Estimation}
\label{algo:update_rules}
\end{algorithm}


In Algorithm~\ref{algo:update_rules}, $\beta_0$ and $\beta_n$ represent the initial and end state (value) of the data attributes modified by an activity. For example, consider an attribute representing inventory levels. Linear regression ($UR_1$) can predict linear relationships from the initial states, such as increasing inventory linearly with each delivery by training a model with transitions, passing $\beta_0$ (initial state) and $\beta_n$ (end state) as input. Decision tree regression ($UR_2$) models more complex data patterns by splitting them into segments based on specific criteria, such as different supplier behaviors, based on $\beta_0$ to predict $\beta_n$. Distribution-based update rules ($UR_3$) model changes using the relationship $d_{i+1} = d_i + \Delta$, where $\Delta$ is drawn from a probabilistic distribution, capturing the variability in changes by performing curve fitting on $\beta_n - \beta_0$. Distribution generators ($UR_4$) produce values based only on each end state $\beta_n$, reflecting the overall distribution of attribute values without considering the initial state. A Markov matrix ($UR_5$) models state transitions based on the current state for categorical attributes. Probabilistic generators ($UR_6$) predict the next state based on observed frequencies independent of the current value. These methods are well-established in predictive modeling and process mining for their efficacy in capturing linear and complex patterns~\cite{Aalst16}.

We evaluate these methods by comparing prediction errors for each activity-attribute pair under global and event-level assumptions, selecting the one with the minimum error. Only activities for which update rules are discovered appear in the mapping $MUR$ of the discovered model (cf., Definition~\ref{def:das}).

\subsection{Branching Conditions Extraction}

To estimate branching conditions at decision points within the process model, our approach captures the data states before and after decision points, such as {\it XOR} and {\it OR} split gateways, and analyzes these data states to determine the conditions influencing the flow paths. To this end, it uses a replayer to simulate the process execution based on an event log $L$ over a process model $P$ written in the BPMN notation.



The approach initializes an empty state matrix $S$. For each process case $T$ in the event log $L$, the initial data state is set up from the first recorded data attribute. The replayer simulates the process execution, tracking changes in data attributes across each event. At decision points, it captures the data states before and after traversing gateways, storing and associating them with the outgoing flows in $S$.

Then, the approach builds a decision tree using state matrix $S$. One decision tree is built for each conditional flow $F$. Every event in the log leading to the gateway $G$, which is the source of $F$, is used in the training sample. The features of this training sample are the attribute states after each (incoming) event completion. The label is $True$ if the flow was traversed right after this state, and it is $False$ otherwise. From the decision tree, we extract a branching condition for that flow (cf., $BC$ in Definition~\ref{def:das}). In addition to assigning branching conditions, the approach also assigns branching probabilities (cf., $BC$ in Definition~\ref{def:das}) to each conditional flow based on the traversal frequencies of each gateway (normalized so that these values are between zero and one). 



\section{Implementation and Evaluation}
\label{sec:evaluation}

The approach was implemented in the {\sc Simod} tool for automated discovery of simulation models from event logs~\cite{CamargoDG20}\footnote{\url{https://github.com/AutomatedProcessImprovement/Simod}} and the {\sc Prosimos} simulation engine~\cite{Lopez-PintadoDB24, Lopez-PintadoHD22}\footnote{\url{https://github.com/AutomatedProcessImprovement/Prosimos}} to support data-aware simulation. Using these extended versions of {\sc Simod} and {\sc Prosimos}, we conducted an empirical evaluation to address the following questions: {\bf EQ1} How accurately are global, event, and case attributes classified when discovered from event logs? {\bf EQ2} How accurately are the update rules identified and mapped to activities when discovered from event logs? and {\bf EQ3} How do branching conditions discovered from event logs and embedded in the data-aware simulation perform relative to non-data-aware (probabilistic) simulation models?

 \subsection{Datasets and Experimental Setup}

We generated 150 synthetic event logs and used three real-life logs to evaluate accuracy in attribute discovery and control flow replication. The main experimental setup is briefly outlined below; detailed procedures are available in the supplementary material at~\url{https://zenodo.org/records/13341463}.


To answer the first two experimental questions {\bf EQ1} and {\bf EQ2}, a synthetic log was generated using the Apromore platform,\footnote{\url{https://apromore.com}} starting from a loan application process simulation model that already incorporated control flow, organizational, and temporal perspectives. We then injected various data attributes into the log generated from this model to assess whether our technique could accurately discover them. We used various data generation functions commonly found in real-life phenomena to generate data attributes (see supplementary material for a detailed explanation).

For numeric (continuous) data patterns, we employed 10 functions: linear trend ({\bf LT}), exponential growth ({\bf EG}), lognormal data ({\bf LN}), autoregressive sequences ({\bf AR1}), conditional exponential growth ({\bf CE}), periodic (sinusoidal) oscillations ({\bf SS}), switching regression ({\bf SR}), piecewise linear trends ({\bf PL}), uniform distribution ({\bf UD}), and normal distribution ({\bf ND}). Each function was applied to generate attributes under four different data patterns: event attributes modified by one (randomly selected) activity ({\bf SE}), event attributes modified by all activities ({\bf ME}), global attributes modified by one (randomly selected) activity ({\bf SG}), and global attributes modified by all activities ({\bf MG}). To emulate real scenarios, each generated sequence was perturbed with different levels of Gaussian noise. This setup resulted in 40 distinct datasets.

 For categorical (discrete) data, the evaluation included 10 patterns modeling state transitions and random state sequences: high self-transition ({\bf HST}), uniform exclusive transitions ({\bf UET}), uniform inclusive transitions ({\bf UIT}), rare state transition ({\bf RST}),  one-state dominating transitions ({\bf SDT}), cyclic transitions ({\bf CT}), and random states with two equal probabilities ({\bf E2P}), two non-equal probabilities ({\bf N2P}), five equal probabilities ({\bf E5P}), and five non-equal probabilities ({\bf N5P}). For example, high self-transition probability and rare transitions to one specific state simulated different Markov chains, while random state functions generated sequences of categorical states based on specified probabilities. These patterns were applied under the same four data patterns as the numeric functions ({\bf SE}, {\bf ME}, {\bf SG}, {\bf MG}), resulting in 40 additional data patterns.

 For case attributes, which are more straightforward to discover, we used 10 data patterns. These included fixed values and attributes generated from exponential, normal, and uniform distributions. We also included 6 categorical data patterns with 2, 5, or 10 states, each configured with equal and non-equal increasing probabilities. 

 We focused on control flow to assess {\bf EQ3}, as branching conditions impact the order in which activities are executed. We used a data-unaware simulation model ({\sc NDAS}) with probabilistic gateway decisions~\cite{Lopez-PintadoDB24} as a baseline to compare with our data-aware model ({\sc DAS}). The BPMN model included three sequential blocks, each with an activity, a split gateway, five outgoing activities, and a join gateway. We tested two BPMN variants: inclusive (OR) and exclusive (XOR) gateways. We allocated 100 full-day resources to avoid non-data-related waiting times and assigned random distribution functions to each activity's processing times. Both {\sc NDAS} and {\sc DAS} models used the same components for fair comparison, differing only in branching decision models. {\sc DAS} used the branching conditions from this paper, while {\sc NDAS} used branching probabilities from frequency analysis. We generated 60 simulated event logs using {\sc Prosimos}.

 We used patterns based on categorical and continuous attributes to evaluate XOR conditions. Categorical patterns included: ``equal" ({\bf EQ}), where all outgoing arcs had conditions holding with the same probability; ``unbalanced" ({\bf  UB}), in which only one flow arc had a 100\% valid condition; and ``random" ({\bf RD}), i.e., arcs had conditions enabled with different probabilities. Continuous patterns followed ``normal" ({\bf  ND}) and ``exponential" ({\bf  ED}) distributions using the $\in$ interval operator, with varying probabilities. Additionally, we used five ``complex conditions" ({\bf  CC1}, {\bf CC2}, {\bf  CC3}, {\bf  CC4}, and {\bf  CC5}) combining categorical and continuous attributes with arithmetic ($=$, $<$, $\leq$, $\in$) and logical ($\land$, $\lor$) operators. Additionally, each pattern was randomly perturbated, adding 20\% noise, i.e., data values not filling any condition. This setup resulted in 20 datasets, each simulated first assuming event-based attributes and then case-based attributes, retrieving 40 event logs.

 Unlike XOR gateways, OR gateways select multiple concurrent flows based on various conditions being fulfilled. To assess this behavior, we used the same data patterns as for XOR ({\bf EQ}, {\bf UB}, {\bf ND}, {\bf ED}, and {\bf CC1}) but extended them to allow the activation of multiple arcs. For each pattern, we generated conditions to activate 1, 2, and 5 random outgoing flow arcs. Additionally, we included an extra case, injecting 20\% noise where none of the conditions held. This setup resulted in a total of 20 event logs.

 In addition to the 150 synthetic datasets,\footnote{Each synthetic event log consisted of 2000 cases, with a minimum of 20000 events varying based on the data patterns generator.} 90 for attribute and update rules accuracy, and 60 for branching conditions, we assessed our approach on three real-life logs. The first two, {\bf Traffic} and {\bf Sepsis}, were also used to evaluate the decision-mining technique presented in~\cite{MannhardtLRA16}. The road fines ({\bf Traffic}) log comes from an Italian local police information system handling traffic fines. The {\bf Sepsis} log records patient pathways with suspected sepsis, a life-threatening infection, over one year in a hospital. The third real-log, {\bf BPIC19},\footnote{\url{https://doi.org/10.4121/uuid:d06aff4b-79f0-45e6-8ec8-e19730c248f1}} comes from a multinational coatings and paints company in the Netherlands, describing the purchase order handling process for its 60 subsidiaries. From this log, we evaluated all the cases covering over 2.5 months of data from January to March 2018. Table~\ref{tbl:log-description} gives descriptive statistics of these real-life business processes.

\begin{table}[tp]
\centering\scriptsize
\caption{Characteristics of the real-life business processes.}\label{tbl:log-description}
\setlength{\tabcolsep}{3.0pt}
\begin{tabular}{lcccccc} \hline
{\bf Log}            & {\bf Cases} & {\bf Events} & {\bf Activities} & {\bf XOR} & {\bf OR}  & {\bf Attributes} \\ \hline
{\bf Trafic}   & 36883 & 130359 & 11         & 8   & 0   & 11 \\ \hline
{\bf Sepsis}   & 1049  & 15214  & 16         & 12  & 0   & 27 \\ \hline
{\bf BPIC19}   & 60740 & 272507 & 37         & 41  & 0   & 12 \\ \hline
\end{tabular}
\end{table}

 To prevent data leakage and overfitting, we split each log into training and testing sets using a temporal split: the first 50\% of traces chronologically for training (to discover the DAS) and the remaining 50\% for testing model accuracy.

 To assess the accuracy of our approach, we used different metrics. We used the Earth Mover's Distance (EMD) for numerical attributes to measure how different two distributions are by calculating the cost to transform one into another, thus fitting for capturing differences in continuous data patterns. For categorical attributes, we applied the Kolmogorov-Smirnov (KS) test, a non-parametric metric that compares the cumulative distributions of two datasets to determine if they differ significantly. The KS test is suitable for categorical data because it is sensitive to differences in the location and shape of the empirical distribution functions. Finally, to assess branching conditions, we used the 3-gram metric to compare the sequences of activities generated by the {\sc DAS} versus the {\sc NDAS} model, i.e., checking how well the {\sc DAS} replicates the process control flow~\cite{Canonne2020, Chapela-CampaBBDKS22}. 

 \subsection{Experimental Results Discussion}

To answer {\bf EQ1}, Table~\ref{tbl:emd_numeric} shows the EMD results for the discovered update rules across 40 synthetic continuous datasets. It highlights correctly classified attributes (global or event-based) in green and misclassified ones in red. Superscripts indicate the model type giving the best approximation: L for linear regression, T for decision tree regression, E for distribution-based rules, and G for generators. Overall, the models led to accurate classifications for approximately 95\% of the attributes, showing their effectiveness in correctly identifying event-based and global modifications. Linear regression was the best model in about 40\% of cases, particularly for linear trends and some exponential growth scenarios. Decision tree regression performed best in 37.5\% of cases, especially with complex, nonlinear data patterns. Distribution-based rules (15\%) and generators (7.5\%) were less frequently the best, indicating their more specialized use cases. 

 \begin{table}[tp]
\centering\scriptsize
\caption{Mean EMD results for synthetic numeric update rules}\label{tbl:emd_numeric}

\setlength{\tabcolsep}{1.0pt}

\begin{tabular}{c|cccccccccc}
\hline
 & {\bf LT} & {\bf EG} & {\bf LN} & {\bf AR1} & {\bf CE} & {\bf SS} & {\bf SR} & {\bf PL} & {\bf UD} & {\bf ND} \\
\hline
{\bf SE} & \cellcolor{lightgreen}$0.16^{L}$ & \cellcolor{lightgreen}$0.15^{L}$ & \cellcolor{lightgreen}$0.05^{T}$ & \cellcolor{lightgreen}$0.04^{L}$ & \cellcolor{lightgreen}$0.03^{T}$ & \cellcolor{lightred}$0.00^{T}$ & \cellcolor{lightgreen}$0.15^{T}$ & \cellcolor{lightgreen}$0.16^{L}$ & \cellcolor{lightgreen}$2.51^{L}$ & \cellcolor{lightgreen}$10.42^{G}$ \\ \hline
{\bf ME} & \cellcolor{lightgreen}$0.05^{L}$ & \cellcolor{lightgreen}$0.09^{L}$ & \cellcolor{lightgreen}$0.05^{L}$ & \cellcolor{lightgreen}$0.02^{T}$ & \cellcolor{lightgreen}$0.04^{L}$ & \cellcolor{lightgreen}$0.12^{T}$ & \cellcolor{lightgreen}$0.27^{T}$ & \cellcolor{lightgreen}$1.37^{T}$ & \cellcolor{lightgreen}$2.35^{T}$ & \cellcolor{lightred}$7.37^{G}$ \\ \hline
{\bf SG} & \cellcolor{lightgreen}$0.04^{L}$ & \cellcolor{lightgreen}$7M^{L}$ & \cellcolor{lightgreen}$0.01^{L}$ & \cellcolor{lightgreen}$0.00^{L}$ & \cellcolor{lightgreen}$8.8K^{T}$ & \cellcolor{lightgreen}$0.01^{L}$ & \cellcolor{lightgreen}$0.09^{\Delta}$ & \cellcolor{lightgreen}$0.28^{\Delta}$ & \cellcolor{lightgreen}$0.19^{T}$ & \cellcolor{lightgreen}$5.37^{T}$ \\ \hline
{\bf MG} & \cellcolor{lightgreen}$1.04^{T}$ & \cellcolor{lightgreen}$8E34^{\Delta}$ & \cellcolor{lightgreen}$0.03^{L}$ & \cellcolor{lightgreen}$0.02^{L}$ & \cellcolor{lightgreen}$3E35^{\Delta}$ & \cellcolor{lightgreen}$0.23^{T}$ & \cellcolor{lightgreen}$1.42^{\Delta}$ & \cellcolor{lightgreen}$6.81^{\Delta}$ & \cellcolor{lightgreen}$2.32^{T}$ & \cellcolor{lightgreen}$7.85^{G}$ \\ \hline
\end{tabular}
\end{table}


 Regarding {\bf EQ2}, update rules for event attributes applied when attributes were modified by single or multiple activities ({\bf SE}, {\bf ME}) were generally well-approximated, with linear and decision tree regressions performing satisfactorily. The EMD values for these cases shown in Table~\ref{tbl:emd_numeric} were relatively low, considering it is a global metric whose value ranges vary from the dataset, indicating a good approximation when applied in event attribute scope. Update rules over global attributes scope, especially those modified by multiple events ({\bf MG}), posed more challenges but resulted in reasonably low EMD values in many cases. The difficulty increases with the accumulated effects of various events, yet the models often maintain a good approximation level. The only two incorrect approximations, i.e., the worst evaluations, were in scenarios involving exponential growth ({\bf EG}) and conditional exponential growth ({\bf CE}) for global attributes ({\bf SG} and {\bf MG}). These resulted in significantly high EMD values due to system overflow caused by the fast growth.  

 Table~\ref{tbl:categorical} shows the KS results for the discovered update rules across 40 synthetic categorical datasets using Markovian models (M) and probabilistic generators (F) as qualifiers. Correct attribute classifications are in green; misclassifications are in red, and cases where the KS value was identical for both event-based and global classifications are in blue. The superscript ``/" indicates results were the same for Markovian models and probabilistic generators, so either could be used. Regarding {\bf EQ1} and {\bf EQ2}, the models performed well in classifying categorical attributes and their update rules, respectively. Misclassifications occurred in 15\% of the cases, primarily in more complex scenarios with multiple update rules modifying the attributes, i.e., {\bf ME} and {\bf MG}. Still, in those cases, differences between global and event-based KS evaluations were very close, and the overall low KS values across most classifications indicate that these misclassifications have a minimal impact on overall estimation accuracy.


\begin{table}[tp]
\centering\scriptsize
\caption{Mean KS results for synthetic categorical update rules}\label{tbl:categorical}
\setlength{\tabcolsep}{1.0pt}
\begin{tabular}{c|cccccccccc} \hline
 & {\bf HST} & {\bf UET} & {\bf RST} & {\bf UIT} & {\bf SDT} & {\bf CT} & {\bf E2P} & {\bf N2P} & {\bf E5P} & {\bf N5P} \\ \hline
{\bf SE} & \cellcolor{lightgreen}$0.33^{/}$ & \cellcolor{lightgreen}$0.25^{/}$ & \cellcolor{lightgreen}$0.25^{/}$ & \cellcolor{lightgreen}$0.33^{/}$ & \cellcolor{lightgreen}$0.25^{/}$ & \cellcolor{lightgreen}$0.20^{F}$ & \cellcolor{lightgreen}$0.33^{/}$ & \cellcolor{lightgreen}$0.33^{/}$ & \cellcolor{lightgreen}$0.17^{F}$ & \cellcolor{lightgreen}$0.17^{/}$ \\ \hline
{\bf ME} & \cellcolor{lightblue}$0.33^{/}$ & \cellcolor{lightgreen}$0.29^{/}$ & \cellcolor{lightgreen}$0.25^{M}$ & \cellcolor{lightred}$0.31^{F}$ & \cellcolor{lightred}$0.26^{F}$ & \cellcolor{lightgreen}$0.02^{M}$ & \cellcolor{lightblue}$0.31^{F}$ & \cellcolor{lightblue}$0.33^{/}$ & \cellcolor{lightgreen}$0.25^{F}$ & \cellcolor{lightgreen}$0.18^{M}$ \\ \hline
{\bf SG} & \cellcolor{lightgreen}$0.02^{M}$ & \cellcolor{lightgreen}$0.25^{M}$ & \cellcolor{lightgreen}$0.22^{M}$ & \cellcolor{lightgreen}$0.33^{/}$ & \cellcolor{lightgreen}$0.25^{M}$ & \cellcolor{lightgreen}$0.20^{M}$ & \cellcolor{lightgreen}$0.30^{M}$ & \cellcolor{lightgreen}$0.23^{M}$ & \cellcolor{lightgreen}$0.24^{/}$ & \cellcolor{lightgreen}$0.17^{M}$ \\ \hline
{\bf MG} & \cellcolor{lightred}$0.31^{M}$ & \cellcolor{lightblue}$0.25^{M}$ & \cellcolor{lightblue}$0.25^{F}$ & \cellcolor{lightblue}$0.33^{/}$ & \cellcolor{lightred}$0.25^{F}$ & \cellcolor{lightgreen}$0.26^{M}$ & \cellcolor{lightred}$0.29^{M}$ & \cellcolor{lightred}$0.31^{M}$ & \cellcolor{lightgreen}$0.25^{M}$ & \cellcolor{lightgreen}$0.18^{/}$ \\
\hline
\end{tabular}
\end{table}

 More specifically, regarding {\bf EQ2}, update rules by probabilistic generators (F) generally provided good approximations for single-event modifications ({\bf SE}), although both Markovian models and generators could equally model eight patterns in {\bf SE}. Both models performed well for multiple-event modifications ({\bf ME}), with some misclassifications in uniform event-based scenarios. Markovian models performed better in global settings with many changes and long-term dependencies, making them more suitable for complex problems requiring more persistent relationships. Markovian models also performed well under single modifications, making them a preferred choice over probabilistic generators in most cases.

 Concluding with {\bf EQ1} and {\bf EQ2}, the 10 synthetic scenarios related to case attributes were classified correctly, showing low EMD and KS values. The full results are available in the reproducibility package.

 In the following, we focus on answering {\bf EQ3}. Table~\ref{tbl:xor_cond} presents the 3-gram results for synthetic XOR gateways across the 40 synthetic scenarios, including event-based and case-based settings, with and without noise. The results assess how well the Data-Aware Model (DAS) replicates the process control flow compared to the Data-Unaware Model (NDAS). Cells highlighted in light blue indicate the lowest values in each column, indicating the best performance.

\begin{table}[tp]
\centering\scriptsize
\setlength{\tabcolsep}{1.0pt}
\caption{3-gram results for synthetic XOR gateways}\label{tbl:xor_cond}
\begin{tabular}{ccc|c|c|c|c|c|c|c|c|c|c}
\hline
\multicolumn{3}{c|}{} & {\bf EQ} & {\bf UB} & {\bf RD} & {\bf ND} & {\bf ED} & {\bf CC1} & {\bf CC2} & {\bf CC3} & {\bf CC4} & {\bf CC5} \\ \hline
\multirow{4}{*}{\rotatebox[origin=c]{90}{{\bf EVENT}}} & \multirow{2}{*}{\rotatebox[origin=c]{90}{{\bf pure}}} & 
    {\bf NDAS} & $0.057$ & \cellcolor{lightblue}$0.000$ & $0.053$ & $0.046$ & $0.042$ & $0.047$ & $0.046$ & \cellcolor{lightblue}$0.047$ & $0.041$ & $0.056$ \\
& & {\bf DAS}  & \cellcolor{lightblue}$0.046$ & \cellcolor{lightblue}$0.000$ & \cellcolor{lightblue}$0.041$ & \cellcolor{lightblue}$0.041$ & \cellcolor{lightblue}$0.040$ & $0.050$ & \cellcolor{lightblue}$0.043$ & \cellcolor{lightblue}$0.047$ & \cellcolor{lightblue}$0.030$ & \cellcolor{lightblue}$0.040$ \\ \cline{4-13}

& \multirow{2}{*}{\rotatebox[origin=c]{90}{{\bf noise}}} & 
    {\bf NDAS} & \cellcolor{lightblue}$0.050$ & \cellcolor{lightblue}$0.030$ & \cellcolor{lightblue}$0.043$ & \cellcolor{lightblue}$0.044$ & \cellcolor{lightblue}$0.035$ & $0.043$ & \cellcolor{lightblue}$0.046$ & $0.047$ & \cellcolor{lightblue}$0.050$ & $\cellcolor{lightblue}0.042$ \\ 
& & {\bf DAS}  & $0.145$ & $0.177$ & $0.138$ & $0.069$ & $0.085$ & \cellcolor{lightblue}$0.041$ & $0.091$ & \cellcolor{lightblue}$0.046$ & $0.105$ & $0.060$ \\ \hline

\multirow{4}{*}{\rotatebox[origin=c]{90}{{\bf CASE}}} & \multirow{2}{*}{\rotatebox[origin=c]{90}{{\bf pure}}} & 
    {\bf NDAS} & $0.205$ & \cellcolor{lightblue}$0.000$ & $0.193$ & $0.195$ & $0.185$ & \cellcolor{lightblue}$0.109$ & $0.201$ & \cellcolor{lightblue}$0.080$ & $0.192$ & $0.196$ \\ 
& & {\bf DAS}  & \cellcolor{lightblue}$0.042$ & \cellcolor{lightblue}$0.000$ & \cellcolor{lightblue}$0.020$ & \cellcolor{lightblue}$0.029$ & \cellcolor{lightblue}$0.031$ & $0.114$ & \cellcolor{lightblue}$0.099$ & $0.081$ & \cellcolor{lightblue}$0.032$ & $\cellcolor{lightblue}0.036$ \\ \cline{4-13}

& \multirow{2}{*}{\rotatebox[origin=c]{90}{{\bf noise}}} & 
    {\bf NDAS} & $0.174$ & \cellcolor{lightblue}$0.076$ & $0.160$ & $0.163$ & $0.156$ & \cellcolor{lightblue}$0.142$ & $0.164$ & \cellcolor{lightblue}$0.089$ & $0.140$ & $0.142$ \\
& & {\bf DAS}  & \cellcolor{lightblue}$0.131$ & $0.140$ & \cellcolor{lightblue}$0.143$ & \cellcolor{lightblue}$0.117$ & \cellcolor{lightblue}$0.084$ & $0.146$ & \cellcolor{lightblue}$0.060$ & $0.092$ & \cellcolor{lightblue}$0.131$ & \cellcolor{lightblue}$0.112$ \\ \hline
\end{tabular}
\end{table}

In scenarios with event-based attributes without noise, both NDAS and DAS performed well, but DAS consistently achieved lower values. In the presence of noise, NDAS performs better, suggesting it is more resilient to noise in event-based scenarios. Yet, the evaluation for XOR gateways shows comparable results when assuming event-based attributes. However, the DAS model outperformed NDAS in settings with case-based attributes, achieving lower values across most scenarios, including noisy and non-noisy ones. Therefore, it indicates the ability of the data-aware model to capture longer-distance associations compared to the data-unaware model. For example, event-based conditions in the DAS relied on values that were updated stochastically before each gateway. On the other hand, the case-based conditions relied on the same attribute value across each gateway, a behavior that probabilistic (NDAS) models cannot replicate as they roll a die at each decision point.

\begin{table}[tp]
\centering\scriptsize
\setlength{\tabcolsep}{1.0pt}
\caption{3-gram results for synthetic OR gateways}\label{tbl:or_cond}

\begin{tabular}{cc|cc|cc|cc|cc|cc}
\hline
\multirow{2}{*}{} & \multirow{2}{*}{} & \multicolumn{2}{c|}{{\bf EQ}} & \multicolumn{2}{c|}{{\bf UB}} & \multicolumn{2}{c|}{{\bf ND}} & \multicolumn{2}{c|}{{\bf ED}} & \multicolumn{2}{c}{{\bf CC}} \\
 & & {\bf NDAS} & {\bf DAS} & {\bf NDAS} & {\bf DAS} & {\bf NDAS} & {\bf DAS} & {\bf NDAS} & {\bf DAS} & {\bf NDAS} & {\bf DAS} \\ \hline
\multirow{4}{*}{}
 & {\bf F1}    & $0.175$ & \cellcolor{lightblue}$0.036$ & \cellcolor{lightblue}$0.000$ & \cellcolor{lightblue}$0.000$ & $0.167$ & \cellcolor{lightblue}$0.048$ & $0.157$ & \cellcolor{lightblue}$0.041$ & $0.172$ & \cellcolor{lightblue}$0.028$ \\ \cline{3-12}
 & {\bf F2}    & $0.279$ & \cellcolor{lightblue}$0.042$ & $0.382$ & \cellcolor{lightblue}$0.015$ & $0.193$ & \cellcolor{lightblue}$0.047$ & $0.177$ & \cellcolor{lightblue}$0.036$ & $0.222$ & \cellcolor{lightblue}$0.055$ \\ \cline{3-12}
 & {\bf F5}    & $0.632$ & \cellcolor{lightblue}$0.053$ & $0.636$ & \cellcolor{lightblue}$0.056$ & $0.637$ & \cellcolor{lightblue}$0.060$ & $0.636$ & \cellcolor{lightblue}$0.056$ & $0.635$ & \cellcolor{lightblue}$0.051$ \\ \cline{3-12}
 & {\bf Noise} & $0.134$ & \cellcolor{lightblue}$0.048$ & \cellcolor{lightblue}$0.091$ & $0.141$ & $0.146$ & \cellcolor{lightblue}$0.069$ & $0.136$ & \cellcolor{lightblue}$0.067$ & $0.149$ & \cellcolor{lightblue}$0.059$ \\ \hline
\end{tabular}
\end{table}



Table~\ref{tbl:or_cond} presents the 3-gram results for synthetic OR gateways across 20 synthetic scenarios, including event-based, in which 1, 2, and 5 flow arcs are activated, and a last scenario with 20\% of added noise. Here, the DAS consistently beats the NDAS, achieving lower values in all measured categories. The DAS superiority is more evident in scenarios with multiple paths and data dependencies, e.g., five flow paths enabled simultaneously (F5). The performance differences between DAS and NDAS are bigger in OR than in XOR gateways. XOR gateways are simpler, directing the flow to one path, which allows NDAS to perform relatively well. However, in OR gateways, the randomness of choosing multiple paths (like rolling a die) plays against NDAS. DAS handles these complexities and data-driven decisions better, giving it an edge in accurately replicating control flow.

Finally, Table~\ref{tbl:real_logs} compares the performance of the DAS and NDAS across three real-life logs using the 3-gram metric and additionally shows the types of attributes discovered by the DAS. Although DAS performed better than NDAS in the 3 logs, the differences in the metric were minor, indicating comparable results. These findings align with the synthetic evaluation results, where NDAS performed relatively well with XOR gateways, the only type present in the event logs. The (slightly) bigger difference was observed in the BPIC19 log, where DAS discovered only case-based attributes, consistent with synthetic evaluations showing DAS's strength in such scenarios. While the specific attribute types in the real-life logs are unknown, the models' accurate replication of control flow suggests that the classification aligns with the process behavior. Full results, including EMD and KS metrics, are available in the reproducibility package.

\begin{table}[tp]
\centering\scriptsize
\setlength{\tabcolsep}{6pt}
\renewcommand{\arraystretch}{1.5}
\caption{Real-Life Logs Results}\label{tbl:real_logs}
\begin{tabular}{c|cc|ccc}
\hline
{\bf Log} & {\bf NDAS} & {\bf DAS} & {\bf Case} & {\bf Global} & {\bf Event} \\ \hline
{\bf Trafic} & $0.763$ & \cellcolor{lightblue}$0.751$ & $1$ & $0$ & $7$ \\ \hline
{\bf Sepsis} & $0.813$ & \cellcolor{lightblue}$0.795$ & $1$ & $3$ & $23$ \\ \hline
{\bf BPIC19} & $0.953$ & \cellcolor{lightblue}$0.927$ & $12$ & $0$ & $0$ \\ \hline
\end{tabular}
\end{table}


\smallskip\noindent\textbf{Reproducibility.} The source code, datasets, models, and instructions to reproduce the experiments can be found at:  \url{https://github.com/orlenyslp/DAS-RP}.
This repository also includes additional details on the experiment setup and results.

\section{Conclusions and Future Work}
\label{sec:conclusion}

This paper introduced a data-aware BPS modeling approach and a method to discover such models from event logs. The approach supports three types of data attributes (global, case-level, and event-level) and incorporates deterministic and stochastic attribute update rules and data-aware branching conditions. The experimental evaluation shows that the techniques for discovering update rules capture a range of common functions and that, in the context of real-life logs, the discovered data-aware models more closely replicate the observed traces relative to data-unaware simulation models.

The proposed approach  discovers rules where the new value of an attribute, after a given activity is executed, depends only on this attribute's previous value. In future work, we plan to extend the approach to discover update rules where the updated value of an attribute is a function over multiple attributes.

Meanwhile, in the proposed BPS, data attributes determine the behavior of the process at the control-flow level only (via branching conditions on the decision gateways). Another future work direction is to extend the proposed BPS modeling and discovery approach to capture situations where the start time or duration of activities may depend on the data attributes. For example, in a loan origination process, the duration of an activity ``Assess credit risk'' may depend on attributes such as the type of loan and the requested loan amount.


\bibliographystyle{IEEEtran}
\bibliography{references}

\end{document}